\documentclass[pmlr]{jmlr}

\usepackage{longtable}
\usepackage{courier}
\usepackage{booktabs}
\usepackage[load-configurations=version-1]{siunitx} 
\usepackage{pbox}

\newcolumntype{L}{>{\centering\arraybackslash}m{2cm}} 
\newcommand\btrule[1]{\specialrule{#1}{0pt}{0pt}}

 \firstpageno{1}

\title{Learning Clinical Outcomes from Heterogeneous Genomic Data Sources}
 \author{\Name{Safoora Yousefi}          \Email{safoora.yousefi@emory.edu} \\
       \addr Department of Computer Science, Emory University\\
       Atlanta, GA, USA     
       \AND
       \Name{Amirreza Shaban}          \Email{amirreza@gatech.edu} \\
       \addr College of Computing, Georgia Institute of Technology\\
       Atlanta, GA, USA 
       \AND
       \Name{Mohamed Amgad}          \Email{mtageld@emory.edu} \\
       \addr Department of Biomedical Informatics, Emory University School of Medicine\\
       Atlanta, GA, USA 
      \AND
      \Name{Ramraj Chandradevan}          \Email{cramraj8@gmail.com} \\
      \addr Department of Electronic and Telecommunication Engineering, University of Moratuwa\\
      Moratuwa, Sri Lanka
      \AND
      \Name{Lee A. D. Cooper} \Email{lee.cooper@emory.edu} \\
      \addr Department of Biomedical Informatics, Emory University School of Medicine\\
      Atlanta, GA, USA}

\begin{document}

\maketitle

\begin{abstract}

Translating the vast data generated by genomic platforms into reliable predictions of clinical outcomes remains a critical challenge in realizing the promise of genomic medicine largely due to small number of independent samples. In this paper, we show that neural networks can be trained to predict clinical outcomes using heterogeneous genomic data sources via multi-task learning and adversarial representation learning, allowing one to combine multiple cohorts and outcomes in training. We compare our proposed method to two baselines and demonstrate that it can be used to help mitigate the data scarcity and clinical outcome censorship in cancer genomics learning problems.
\end{abstract}

\section{Introduction}
Since the emergence of high throughput experiments such as Next Generation Sequencing, the volume of genomic data produced has been increasing exponentially \citep{stephens2015big}. A single biopsy can generate tens of thousands of transcriptomic, proteomic, or epigenetic features. The ability to generate genomic data has far outpaced the ability to translate these data into clinically-actionable information, as typically only a handful of molecular features are used in diagnostics or in determining prognosis \citep{bailey2018comprehensive,van2002gene,cancer2015comprehensive}. 

Machine-learning has emerged as a powerful tool for analyzing high-dimensional data, with open software tools that enable scalable and distributed data analysis. A sub-field of machine learning, known as deep learning, has recently achieved remarkable success in learning from high dimensional images and sequences \citep{lecun2015deep}. It involves artificial neural networks with several processing layers that learn representations of data with multiple abstraction levels.

There are several challenges in applying neural networks to genomic data \citep{min2017deep}. The more parameters a machine learning model has, the more independent samples it requires for training \citep{abu1989vapnik}, and neural networks often have many thousands of parameters due to their layered nature. Cancer genomic datasets often have small sample size (only hundreds of samples), and much larger dimensionality (tens of thousands of features). Several approaches have been employed to alleviate this data insufficiency including dimensionality reduction, feature selection, data augmentation, and transfer learning \citep{ching2018opportunities}.

An alternative approach is to integrate genomic data from multiple studies and hospitals (for example, two independent studies of breast cancer genomics) to increase training set size. Heterogeneity of available genomic datasets due to technical and sample biases poses challenges to integrating multiple data sources. Cohorts from multiple sources typically have difference demographic or disease stage distributions, may be subject to different signal capture calibration, post-processing artifacts, and naming conventions. This means that naively combining heterogeneous cohorts is both difficult and may degrade model accuracy due to batch effects \citep{tom2017identifying}.

In addition to integrating data from studies involving the same primary cancer site, we may benefit from pooling cohorts diagnosed with different cancer types together to increase training size. Cancers that originate from different primary sites are known to have large differences in genetic markup, although there are some remarkable similarities that seem to play a fundamental role in carcinogenesis \citep{hoadley2018cell, bailey2018comprehensive, hanahan2011hallmarks}. The idea of combining multiple cancer types relies on the premise that models of sufficient complexity and constraints can exploit these similarities to improve outcome prediction.

In this paper we propose a multi-task learning and adversarial representation learning approach that allows integration of heterogeneous cohorts. Experiments demonstrate that our proposed methods can be used to alleviate the data scarcity issue in several cancer genomics learning problems.

\paragraph{Technical Significance}
Building upon SurvivalNet \citep{yousefi2016learning, yousefi2017predicting} -a neural network model for survival prediction- we propose a multi-task learning approach that enables: a) training SurvivalNet on multiple heterogeneous data sources while avoiding the issues that arise from naively combining datasets, and b) training on multiple clinical outcomes from the same cohort, thus helping to address the issue of censorship (loss to follow-up) often encountered in clinical datasets. We further enhance our proposed method by introducing an adversarial cohort classification loss that prevents the model from learning cohort-specific noise, thus enabling task-invariant representation learning.

\paragraph{Clinical Relevance}
Accurate prognostication is crucial to clinical decision making in cancer treatment. Although contemporary sequencing platforms can provide tens to hundreds of thousands of features describing the molecular profiles of neoplastic cells, only a small number of these features have established clinical significance and are used in prognostication. Using machine learning to make reliable and accurate predictions of clinical outcomes from high-dimensional molecular data is a major step in realizing the promise of precision medicine, but has been challenged so far by sample size limitations. This paper aims to address this data insufficiency challenge using techniques that allow training of neural networks on multiple heterogeneous cohorts. The prospect of increasing the training set size by integrating cohorts from different hospitals or studies, and exploiting traits shared among all human cancer types, to train powerful predictive models motivates the task-invariant representation learning approach proposed in this paper.

\section{Background and Related Work}
\subsection{Survival analysis with Cox proportional hazards model}
Survival analysis refers to any problem where the variable of interest is time to some event, which in cancer is often death or progression of disease. Time-to-event modelling is different from ordinary regression due to a specific type of missing data problem known as censoring. Incomplete or censored observations are important to incorporate into the model since they could provide critical information about long-term survivors \citep{harrell2015regression}. The most widely used approach to survival analysis is the semi-parametric Cox proportional hazards model \citep{david1972regression}. It models the hazard function at time $s$ given the predictors $x_i$ of the $i$th sample as:
\begin{equation}
\label{cox-haz}
	h(s|x_i, \beta) = h_0(s)e^{\beta^\top x_i }
\end{equation}
The model parameters $\beta$ are estimated by minimizing Cox's negative partial log-likelihood:
\begin{equation}
\label{cox-loss}
	L_{cox}(X, Y, \beta) = - \sum_{x_i \in U}{\Big(\beta^\top x_i  - \log\sum_{j\in{R_i}}{e^{\beta^\top x_j}}\Big)}
\end{equation}
where $X=\{x_1, ...,x_N\}$ are the samples, and $Y = \{E, S\}$ represents label vectors of event or last follow-up times $S=\{s_1, ...,s_N\}$ and event status $E=\{e_1, ...,e_N\}$. For censored samples ($e = 0$), $s$ represents time of last follow-up while for observed samples ($e = 1$), it represents event time. The outer sum is over the set of uncensored samples $U$ and $R_i$ is the set of \textit{at-risk} samples with $s_j \geq s_i$. The baseline hazard $h_0(t)$ is cancelled out of the likelihood and can remain unspecified. 

We will look at two variants of Cox's model that are used in this paper as baselines. \cite{park2007l1} proposed Cox-ElasticNet which minimizes Cox's loss subject to the \textit{elasticnet} regularization constraint:
\begin{equation}
\label{cox-net}
	L_{CEN}(X, Y, \beta) = L_{cox}(X, Y, \beta) + \gamma(\alpha\|\beta\|_1 + (1-\alpha)\|\beta\|_2^2)
\end{equation}

where $\alpha\in[0,1]$ is the mixture coefficient of $\ell1$ and $\ell2$ regularization terms, and $\gamma\in[0, +\inf)$ is the overall regularization rate.

A non-linear alternative to Cox regression is SurvivalNet \citep{yousefi2016learning,yousefi2017predicting}, a fully connected artificial neural network $f_W$ with parameters $W$ that replaces $X$ in Equation \ref{cox-loss} with its non-linear transformation $f_W(X)$. SurvivalNet has been shown to outperform other common survival analysis techniques such as random survival forests \citep{ishwaran2008random} and Cox-ElasticNet.

This paper proposes two multi-task learning models built upon SurvivalNet to learn task-invariant representations from heterogeneous data sources. These models will be discussed in detail in section \ref{sec:methods}.

\subsection{Multi-task learning for survival analysis}
Both theoretical and empirical studies show that learning multiple related tasks simultaneously often significantly improves performance relative to learning each task independently \citep{baxter2000model, ben2003exploiting, caruana1997multitask}. This is particularly the case when only a few samples per task are available, since with multi-task learning, each task has more data to learn from. Multi-task learning has been applied to many areas of machine learning inlcuding computer vision \citep{zhang2014facial}, natural language processing \citep{collobert2008unified}, and survival analysis \citep{wang2017multi,li2016transfer}.


Following \cite{pan2010survey}, we provide a classification of multi-task learning problem settings in cancer survival analysis. Let us first define the terms \textit{domain} and \textit{task}. A domain is a pair ${\{\mathcal{X}, P(X)\}}$ which includes a feature space and a marginal probability distribution where $X = \{x_1,...,x_n\} \in \mathcal{X}$. A task ${\{\mathcal{Y}, P(Y|X)\}}$ consists of a label space and a conditional probability distribution function. $P(Y|X)$ is the ultimate predictive function that is not observed but can be learned from training data. Multi-task learning, by definition, involves different tasks, that is different $P(Y|X)$, or even different label spaces. With that in mind, we classify multi-task survival analysis problems as follows:

\begin{itemize}
    \item[\bf 1.] Different $P(X)$: Data for the tasks come from different distributions. Examples include: 
        \begin{itemize}
            \item[\bf -] Standard gene expression data and progression-free survival labels are available for all cohorts, but the cohorts are diagnosed with different cancer types.
            \item[\bf -] Standard gene expression data and progression-free survival labels are available for all cohorts, and the cohorts are diagnosed with the same cancer types but belong to different studies/hospitals.
        \end{itemize}
    \item[\bf 2] Different $\mathcal{X}$: Data for the tasks come from different feature spaces. Note that this automatically leads to different $P(X)$. Example of different feature spaces are gene expression data and mutation data.
    \item[\bf 3.] Different $P(Y|X)$: All tasks are the same in nature, but the conditional distribution of labels are different. For example, learning overall survival and progression-free survival simultaneously for the same cohort of patients falls under this category.
    \item[\bf 4] Different $\mathcal{Y}$: This class of multi-task problems involves different prediction tasks (such as survival analysis and classification).
\end{itemize}
    
This paper focuses on scenarios 1 (Sections \ref{sec:tcgamb} and \ref{sec:types}) and 3 (\ref{sec:mtl}). The general form of the loss function when learning $T$ tasks simultaneously is:
\begin{equation}
\label{multi-loss}
	L(Y, X, W) = \sum_{t={1}}^{T} L_t(y^t, g^t(W^t, X^t)) + \gamma\lambda(Y, X, W)
\end{equation}
 $l_t$ and $W^t$, respectively, are the loss function and the parameters of task $t$. $Y=\{Y^1,...,Y^T\}$ and $X=\{X^1,...,X^T\}$  are the combined input data of all $t$ tasks. $g^t$ indicates the prediction function corresponding to task $t$, and $\lambda$ is a regularization or auxiliary function that captures task relatedness assumptions, examples of which include cluster norm \citep{jacob2009clustered}, trace norm \citep{argyriou2007multi}, and $\ell_{1,2}$ norm \citep{liu2009multi}. $\gamma$ is a weight parameter controlling the importance of the auxiliary function.
 
 Previous work has applied multi-task learning under different task relatedness assumptions to train Cox's proportional hazards model using multiple genomic data sources \citep{wang2017multi,li2016transfer}. In this paper, our main assumption is that gene expression data lies on a lower dimensional subspace that can be utilized in several prognostic tasks. We will enforce this assumption via parameter sharing and the bottleneck architecture of our models as shown in Figure \ref{fig:methods}. Moreover, In section \ref{sec:methods} we describe how an adversarial classification objective can be used as auxiliary function $\lambda$ to encourage task-invariant representation learning.
 
For simplicity, we consider settings with one target task and one auxiliary task (T=2). Ground truth labels are available for both tasks, and the goal is to make better predictions on the target task by learning relevant information from the auxiliary task. 

\label{sec:mtl4sa}
\subsection{Adversarial representation learning}
The idea of using adversarial learning to match two distributions was first proposed by \cite{goodfellow2014generative} for training generative models. In generative adversarial models, a generator aims to generate realistic data to mislead a discriminator that is simultaneously trained to distinguish between real and generated data. This competition drives the two components of the model to improve, until the generated data distribution is indistinguishable from the real data distribution. 

This idea has been applied to unsupervised domain adaptation for natural language processing and computer vision, with varying design choices including parameter sharing, type of adversarial loss, and discriminative vs. generative base model \citep{ganin2014unsupervised, ganin2016domain, tzeng2015simultaneous, liu2016coupled, tzeng2017adversarial}.

We adapt this idea to multi-task learning to encourage our proposed model to learn task-invariant representations. A cohort discriminator is trained to assign samples to their cohort. Simultaneously, a SurvivalNet is adversarially trained to confuse the discriminator by learning a representation of data where samples from different cohorts are indistinguishable (in addition to learning to predict survival). This competition will teach SurvivalNet to avoid learning cohort-specific noise. 

\section{Cohort}
The Cancer Genome Atlas (TCGA) provides publicly available clinical and molecular data for 33 cancer types. TCGA gene expression features were taken from the Illumina HiSeq 2000 RNA Sequencing V2 platform. TCGA clinical data contains overall survival (OS) and progression free interval (PFI) labels, with varying degrees of availability for different primary cancer sites \citep{liu2018integrated}. This data has been obtained from multiple hospitals and health-care centers, so a considerable degree of heterogeneity exists within the TCGA. 

PFI is defined as the period from the date of diagnosis until the date of the first occurrence of a new tumor-related event, which includes progression of the disease, locoregional recurrence, distant metastasis, new primary tumor, or death with tumor. OS is the period from the date of diagnosis until the date of death from any cause. Since patients generally suffer from disease progression or recurrence before dying, PFI requires shorter follow-up times and has higher event rate. Additionally, OS is a noisy signal due to deaths from non-cancer causes. Therefore, wherever possible, PFI is used as the outcome variable.

We also used METABRIC (Molecular Taxonomy of Breast Cancer International Consortium) \citep{curtis2012genomic} gene expression and clinical data in section \ref{sec:tcgamb}. Since METABRIC comes with OS labels only, OS was used as the outcome variable in this section.

\subsection{Cohort Selection} 
TCGA breast invasive carcinoma (BRCA) was used in section \ref{sec:tcgamb} as target cohort. In section \ref{sec:types} we perform multi-task learning experiments on every possible pair of target and auxiliary cohorts chosen from a subset of cancer types. Out of the 33 TCGA cancer types, we selected those with PFI event rate higher than 20\%. We used the performance of Cox-ElasticNet \citep{park2007l1} on each of these cancer types as a measure of outcome label quality, and used only those cancer types where Cox-ElasticNet achieved a c-index of 60\% and higher, leaving us with adrenocortical carcinoma (ACC), cervical squamous cell carcinoma (CESC), lower-grade glioma (LGG), kidney renal clear cell carcinoma (KIRC), kidney renal papillary cell carcinoma (KIRP), mesothelioma (MESO), and pancreatic adenocarcinoma (PAAD). ACC and MESO could not be used as target cohorts since their small sample sizes did not allow for reliable model evaluation. All of the mentioned cancer types were used as auxiliary cohorts. 

We discarded samples that did not have gene expression data or outcome labels. A summary of sample sizes and event rates of datasets after this preprocessing step is given in Table \ref{tab:data}. Z-score normalization and 3-NN missing data imputation were performed on gene expression data. No further feature selection or dimensionality reduction was performed. In section \ref{sec:tcgamb}, we found the intersection of Hugo IDs present in both BRCA and METABRIC datasets (17272 genes), and discarded the genes that were absent in either dataset.

\newcolumntype{M}{>{\centering\arraybackslash}m{1.5cm}} 
\begin{table}[tbp]
\centering
\resizebox{.7\textwidth}{!}{%
\begin{tabular}{LMMMM}
\btrule{1pt}
Dataset Name & Number of Samples & Numbr of Features & Event Rate & Event Type \\
\btrule{.5pt}
ACC          & 79        & 20531      & 52\%       & PFI        \\
CESC         & 304       & 20531      & 23\%       & PFI        \\
KIRC         & 533       & 20531      & 30\%       & PFI        \\
KIRP         & 514       & 20531      & 37\%       & PFI        \\
LGG          & 288       & 20531      & 20\%       & PFI        \\
MESO         & 84        & 20531      & 70\%       & PFI        \\
PAAD         & 178       & 20531      & 58\%       & PFI        \\
BRCA         & 1094      & 20531      & 13\%       & OS         \\
METABRIC     & 1903      & 24368      & 33\%       & OS         \\
\hline
\end{tabular}
}
\caption{\small{Summary of datasets.}}
\label{tab:data}
\end{table}
\label{sec:cohort}

\section{Methods}
\begin{figure}[t]
    \centering
    \begin{tabular}{ccc}
 \multicolumn{2}{c}{\includegraphics[width=.8\textwidth]{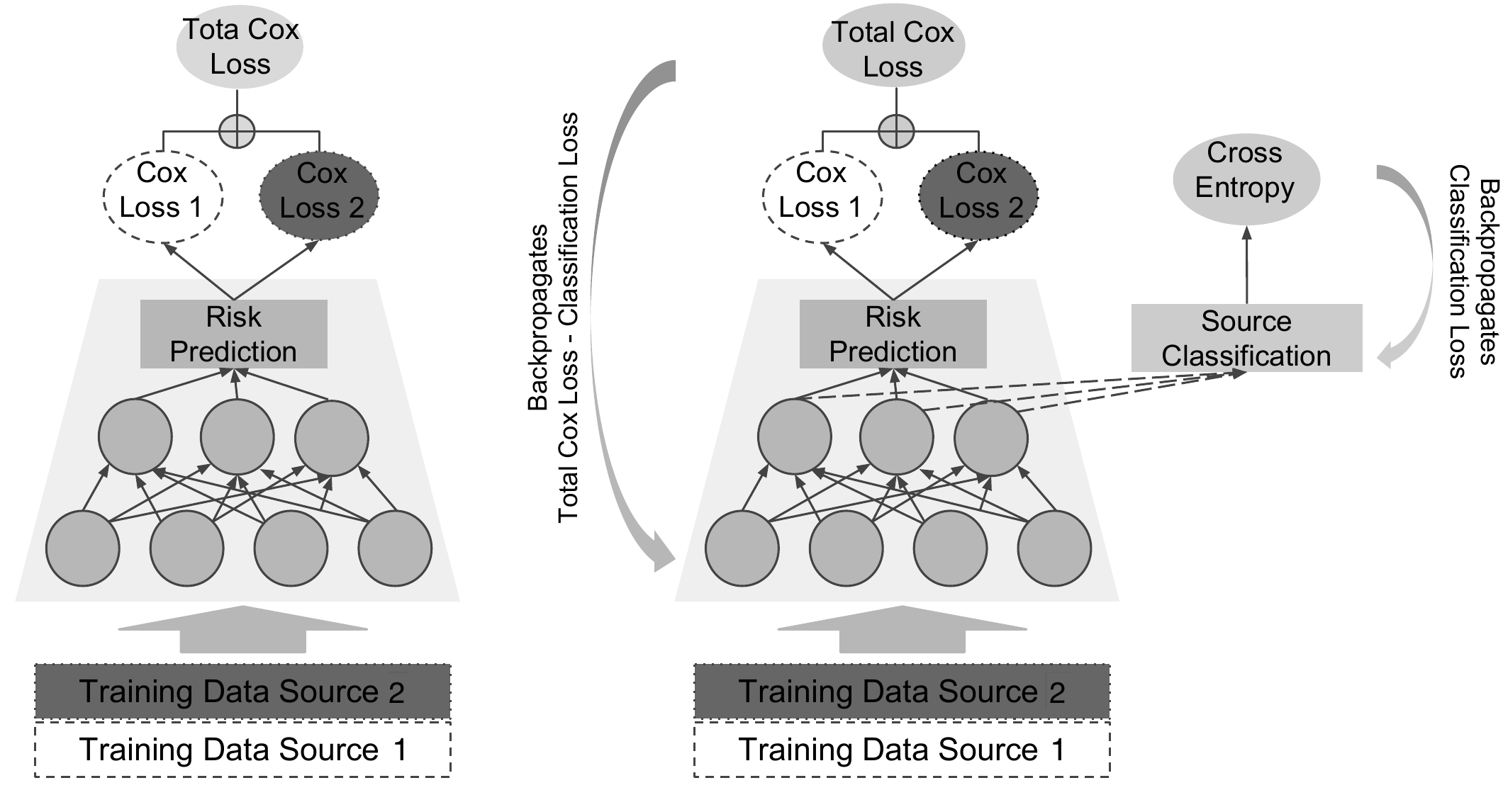}}\\[6pt]
 \footnotesize{a. Multi-task Model (MTL)} & \footnotesize{b. Adversarial Multi-task Model (ADV-MTL)}\\[6pt]
    \end{tabular}
    \caption{\footnotesize{Model architectures used.}}
    \label{fig:methods}
\end{figure}


In cases where the target and auxiliary tasks are similar and their corresponding samples come from similar distributions, a natural approach is to simply combine (i.e. concatenate) the target and auxiliary datasets and train a single-task model on the combined training data, as done is \cite{yousefi2017predicting}. We implement this approach as a baseline using both Cox-ElasticNet and SurvivalNet to provide two performance baselines.

But the assumption that the two cohorts come from the same distribution rarely holds. Comparisons of survival time between pairs of samples are integral to the Cox log-likelihood loss function. When one naively combines datasets to train a model with a single Cox loss, in addition to comparisons within target cohort and within auxiliary cohort, comparisons between these cohorts contribute to the loss. Since the difference between distributions of these cohorts could be due to clinically insignificant factors (such as batch effects), these between-cohort comparisons could be misleading in training. Our first proposed model aims to eliminate this potentially misleading signal from the training process via multi-task learning:

\textbf{Multi-task learning (MTL):} This proposed extension of SurvivalNet model comprises one Cox loss node per each task. Each Cox loss node is responsible for one cohort only, so that only within-cohort comparisons contribute to the loss (See Figure \ref{fig:methods}a). The objective function of the MTL model is the sum of all Cox losses:
    
\begin{equation}
\label{mtl-loss}
	L_{MTL} = \sum_{t=1}^T L_{Cox}(f_W(X^t), Y^t, \beta)
\end{equation}
where $f_W$ is the SurvivalNet model. All parameters of MTL, $\beta$ and $W$, are shared among tasks.

Although we are encouraging sparse representation learning via the bottleneck architecture of the MTL model, that does not adequately force the model to learn a task invariant representation. The model may learn a sparse representation, but still have enough parameters to be able to discriminate between samples from different cohorts and process them differently. The adversarial model described below addresses this limitation. 

\textbf{Adversarial model (ADV):} This models extends SurvivalNet by addition of an adversarial cohort classification loss. Let $X_{comb} = \{x_1, ..., x_M\}$ and $Y_{comb} = \{y_1, ..., y_M\}$ denote the combination of all $X^t$ and $Y^t$, respectively, including $M$ samples in total. A set of one-hot vectors $Y_D = \{d_1, ..., d_M\}$ indicate cohort membership, so that $d_{it}=1$ means that the $i$th sample belongs to the $t$th cohort. A cohort discriminator is trained to assign the transformed samples $z_i = f_W(x_i)$ to the cohort they belong to. This component of the model is a multi-class logistic regression with a softmax cross-entropy loss. It comprises a linear transformation $g_{\theta}$ mapping $z_i$ to a T-dimensional vector, where T is the number of tasks, and a softmax function that transforms the result to a T-dimensional vector of probabilities. The predicted probability that sample $i$ belongs to cohort $t$ is given by:

\begin{align*}
    \hat d_{it} &= \frac{e^{g_{\theta}(z_i)_t}}{\sum_{k=1}^T(e^{g_{\theta}(z_i)_k})},
\end{align*}
and the objective function of the discriminator $L_D$ is the cross-entropy between predicted probabilities and cohort lables:
    \begin{equation}
    \label{adv-loss-D}
     L_D(f_{W}(X_{comb}), Y_D, \theta) = \gamma \sum_{i=1}^{M}\sum_{t=1}^{T} - d_{it} \log{\hat{d_{it}}}
    \end{equation}
  This loss function only trains the parameters of the discriminator, namely $\theta$ the parameters of the linear function $g_{\theta}$. 
 
Simultaneously, a risk predictor is adversarially trained to learn a cohort-invariant representation that misleads the cohort classifier, in addition to learning to predict risk of event. The objective function of the risk predictor component of the model is:
    \begin{equation}
    \label{adv-loss-R}
    	L_R = L_{Cox}(f_{W}(X_{comb}), Y_{comb}, \beta) - \gamma L_D(f_{W}(X_{comb}), Y_D, \theta)
    \end{equation}

$L_R$ trains the parameters of the risk predictor $\beta$ as well as $W$. By updating $W$ with an objective function that is the opposite of that of the discriminator, we encourage learning a representation of data in which samples from different cohorts are indistinguishable. $\gamma$ controls the contribution of the adversarial loss to representation learning.

\textbf{Adversarial multi-task model (ADV-MTL)}: Combining the MTL and ADV model described above, we allocate one Cox loss node for each cohort and additionally employ an adversarial cohort classification (See Figure \ref{fig:methods}b). The discriminator loss function is the same as given by Equation \ref{adv-loss-D} while the rest of the model is trained with the following objective:
    \begin{equation}
    \label{adv-loss-R2}
    	L_R = \sum_{t=1}^T L_{Cox}(f_W(X^t), Y^t, \beta) - \gamma L_D(f_{W}(X_{comb}), Y_D, \theta)
    \end{equation}    
\label{sec:methods}
\section{Results} 

\subsection{Model selection and training} 
We use random stratified sampling to sample $60\%$ of target data as training and use the remaining $40\%$ as hold-out testing data. Stratified sampling ensures similar event rates in training and testing sets. 

Training set is augmented with any auxiliary data at this stage if the experiment calls for it. For model selection, grid search cross validation is performed on the training set using 5 randomly sampled training ($80\%$ of target training data + auxiliary data) and validation sets ($20\%$ of target training data). The selected model is then evaluated on the hold-out testing data. We repeat this procedure on $30$ randomly sampled training and testing sets and use re-sampled t-test and paired re-sampled t-test \citep{dietterich1998approximate} to provide confidence intervals and significance analysis.
 
A single hidden layer with 50 hidden units was used in all neural networks. Learning rate, drop-out regularization rate, and L2 regularization rate of neural network parameters $W$, and the weight of the discriminator loss $\gamma$ were tuned via grid search. In Cox-ElasticNet experiments, we tuned the regularization rate $\gamma$ and the mixture coefficient $\alpha$. 

The same sampling, training, model selection and evaluation procedures was used in all experiments. All software and parameters to reproduce the results presented in this section are publicly available at [GITHUB LINK].


\label{subsec:training}
\subsection{Evaluation Metric}
We measured model performance using \emph{concordance index} (CI) that captures the rank correlation of predicted and actual survival \citep{harrell1982evaluating}, and is given by:
\vspace{-2mm}
\begin{equation}\vspace{-2mm}
    \textit{CI}(\beta, X) =  \sum_P{\tfrac{I(i, j)}{|P|}}
\end{equation}
\begin{equation}
    I(i, j) = 
\begin{cases}
    1,& \text{if } r_j > r_i\ \text{and}\ t_j > t_i\\
    0,              & \text{otherwise}
\end{cases}
\end{equation}

Where $P$ is the set of orderable pairs. A pair of samples ($x_i$, $x_j$) is orderable if either the event is observed for both $x_i$ and $x_j$, or $x_j$ is censored and $t_j > t_i$.
Intuitively, CI measures the pairwise agreement of the prognostic scores $r_i$, $r_j$ predicted by the model and the actual time of event for all orderable pairs. Optimizing the cox partial likelihood (Equation \ref{cox-loss}) has been shown to be equivalent to optimizing CI \citep{steck2008ranking}. 

\subsection{Combining two breast cancer cohorts} 
\begin{figure}
    \centering
    \includegraphics[width=.8\textwidth]{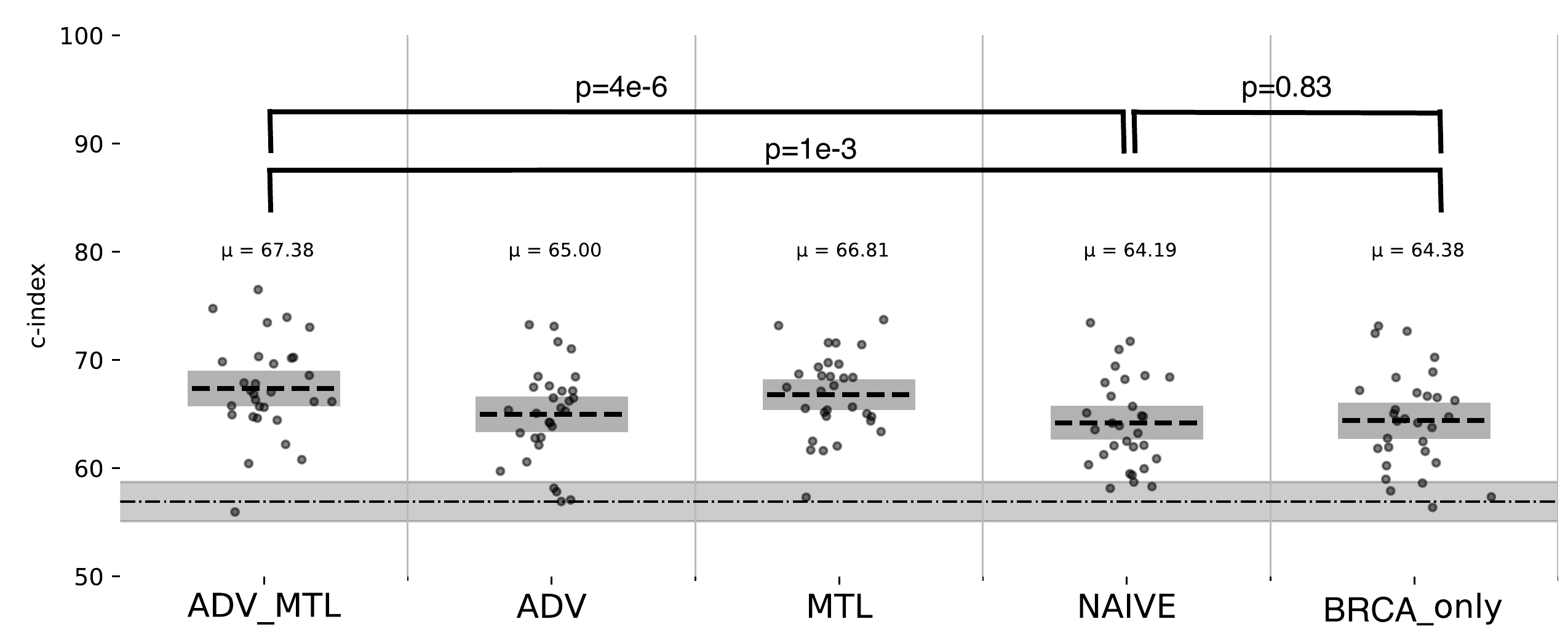}
    \caption{\footnotesize{One solution to data insufficiency is to obtain similar cohorts and combine them with target training data. METABRIC gene expression and clinical data were combined with BRCA training data to improve performance on BRCA using models described in section \ref{sec:methods}. \texttt{BRCA\_only} refers to performance of SurvivalNet when trained on BRCA only. The horizontal line indicates the performance of Cox-Elasticnet trained on naive combination of cohorts. Shaded areas indicate 95\% confidence intervals. Paired t-test p-values are provided for some comparisons.}}
    \label{fig:mbtcga}
\end{figure}

This section investigates the integration of two breast cancer cohorts from independent studies. We use BRCA as target and METABRIC as auxiliary cohort. Both of these cohorts are diagnosed with breast cancer, and have overall survival labels. In such cases where similar biological processes determine the outcomes, one would expect naively pooling cohorts together to lead to better predictions on each of the cohorts. This is the expectation particularly in this case where the auxiliary cohort has twice the number of samples as the target cohort and three times the event rate. We train SurvivalNet and Cox-ElasticNet on the naive combination of BRCA and METABRIC as a two baselines, and compare the results to MTL, ADV, and ADV-MTL models as shown in Figure \ref{fig:mbtcga}. 

Surprisingly, we observe that simply adding METABRIC to training data has no effect on prediction c-index on BRCA (p=0.83). This implies that the distributions of the two datasets are so different that comparisons made between them are not providing useful insights to the model. Therefore, we eliminate these between-cohort comparisons by training a MTL model and observe a significant improvement (p=3e-4). Addition of adversarial classification loss to SurvivalNet lead to a slight improvement over naive SurvivalNet, while combining the MTL and ADV approaches into ADV-MTL significantly outperforms all other methods. The significance of the improvement achieved by ADV-MTL from a machine learning standpoint is discussed in section \ref{sec:lc}. 

\label{sec:tcgamb}

\subsection{Combining cohorts with different cancer types} 
\label{sec:types}
 We repeat the experiments of section \ref{sec:tcgamb} this time using cohorts diagnosed with different cancer types. As explained in section \ref{sec:cohort}, we use five cancer types as target and seven cancer types as auxiliary cohorts, performing experiments on each possible pair of target and auxiliary cohorts. Results of these experiments are summarized in Tables \ref{tab:naive} and \ref{tab:advmtl} in terms of average c-index achieved on target test set. 
 
 Table \ref{tab:naive} summarizes the result of training SurvivalNet on the naive combination of target and auxiliary cohorts (NAIVE). The last column provides c-index of SurvivalNet after training on target cohort only. This naive approach leads to either significant (p\textless0.01) deterioration or no significant difference in performance compared to SurvivalNet trained on target only, achieving significant improvement only in one case. ADV-MTL, on the other hand, achieves significant improvement over target-only setting in 10 cohort combinations (See Table \ref{tab:advmtl}).

\begin{table}
\centering
\resizebox{\textwidth}{!}{%
\begin{tabular}{MMMMMMMM|M}
    \btrule{1pt}

    &  {\bf+ACC}  &  {\bf+CESC} &  {\bf+LGG}  &  {\bf+KIRP} &  {\bf+KIRC} &  {\bf+PAAD} &  {\bf+MESO} &  {\bf Target-only}      \\
\hline
CESC & 63.82 ($\pm1.79$) & -    & 62.28* ($\pm1.91$) & 63.08 ($\pm1.87$) & 61.49* ($\pm1.33$)& 60.50* ($\pm1.63$)& \bf65.43 ($\pm1.44$) & 64.06 ($\pm1.60$) \\
\hline
LGG & \bf71.12 ($\pm0.84$) &  \bf70.87 ($\pm0.96$) & -    & 69.45 ($\pm1.39$) &67.99* ($\pm1.05$)& \bf71.27 ($\pm0.86$)& \bf70.76 ($\pm0.70$)& 70.62 ($\pm0.74$) \\
\hline
KIRP  & \bf75.14 ($\pm1.98$) & 74.17 ($\pm1.93$) & 73.48 ($\pm1.84$) & -    &74.38 ($\pm1.90$)& 74.58 ($\pm1.36$)& \bf74.89 ($\pm1.77$) & 74.62 ($\pm1.83$)\\
\hline
KIRC & \bf71.20 ($\pm1.14$)& \bf70.31 ($\pm0.81$)&  \bf70.62 ($\pm1.08$)& \bf71.03* ($\pm1.16$)& - & \bf70.28 ($\pm1.13$)& 67.32* ($\pm1.36$)& 69.69 ($\pm1.06$)\\
\hline
PAAD &63.77 ($\pm1.43$)& 59.25* ($\pm1.54$)& 63.22 ($\pm1.43$)& 62.24* ($\pm1.81$)& 58.80* ($\pm2.13$)& - & 63.49 ($\pm1.50$)& 63.78 ($\pm1.40$)\\
 \btrule{1pt}

\end{tabular}}
\caption{\footnotesize{Performance of SurvivalNet (c-index) when trained on naive combination of target data (rows) and auxiliary data (columns). Performance after training on target data only has been provided for reference. Numbers is parentheses are 95\% confidence intervals. All improvements are boldened. * marks significant differences (p--values less than 0.01).}}
\label{tab:naive}
\end{table}

\begin{table}
\centering
\resizebox{\textwidth}{!}{
\begin{tabular}{MMMMMMMM|M}
    \btrule{1pt}
    &  {\bf+ACC}  &  {\bf+CESC} &  {\bf+LGG}  &  {\bf+KIRP} &  {\bf+KIRC} &  {\bf+PAAD} &  {\bf+MESO} &  {\bf Target-only}      \\
\hline
CESC & \bf64.32 ($\pm1.75$)& -  & 63.41 ($\pm1.57$)& 61.54 ($\pm1.63$)&  62.92 ($\pm1.84$) & 60.64 ($\pm1.58$) & \bf65.26 ($\pm1.57$) & 64.06 ($\pm1.60$)\\
\hline
LGG & \bf71.04 ($\pm0.93$)&\bf71.80* ($\pm1.05$) & - & \bf71.23* ($\pm0.71$) & \bf71.00 ($\pm0.97$)& \bf72.24* ($\pm0.78$)& 70.42 ($\pm0.74$)& 70.62 ($\pm0.74$)\\
\hline
KIRP  & \bf76.49* ($\pm1.89$)& 72.24* ($\pm1.90$)& \bf74.91 ($\pm1.68$)& -  & \bf75.60 ($\pm1.78$)& \bf75.0 ($\pm1.64$)& \bf75.45 ($\pm1.9$) & 74.62 ($\pm1.83$)\\
\hline
KIRC  & \bf72.59* ($\pm0.92$)& \bf72.18* ($\pm0.85$)& \bf72.50* ($\pm0.74$)& \bf73.14* ($\pm0.75$) & -  & \bf71.83* ($\pm1.0$)& \bf73.53* ($\pm0.94$)& 69.69 ($\pm1.06$)\\
\hline
PAAD  & \bf64.48 ($\pm1.66$)& 62.66* ($\pm1.58$)& \bf64.17 ($\pm1.33$)& \bf63.99 ($\pm1.42$)& \bf64.20 ($\pm1.49$)&  -  &\bf64.14 ($\pm1.66$)& 63.78 ($\pm1.40$)\\

 \btrule{1pt}
\end{tabular}}
\caption{\footnotesize{Performance of ADV-MTL model when trained on combination of target data (rows) and auxiliary data (columns). Performance of SurvivalNet after training on target data only has been provided for reference. Numbers is parentheses are 95\% confidence intervals. All improvements are boldened. * marks significant differences (p--values less than 0.01).}}
\label{tab:advmtl}
\end{table}

\subsection{Combining multiple outcome labels for the same cohort}
\label{sec:mtl}
As discussed in section \ref{sec:cohort}, TCGA samples may have multiple outcome labels. Overall survival (OS) labels are noisier, but simpler to obtain since the patients are either deceased or alive at the end of the study. 
As shown in Table \ref{tab:mtl}, for some patients, a new tumor event is never observed (or recorded) during the study (censored PFI), while their overall survival outcome is observed (deceased by end of study). In such cases, overall survival could provide an extra supervision signal in training a predictive model that originally targets PFI prediction.

We use the MTL model to simultaneously use PFI and OS outcomes in training. Target and auxiliary domains are the same, so there is no need for adversarial domain-invariant representation learning. What differentiates the tasks from each other is the the predictive function $P(Y|X)$. Results are summarized in table \ref{tab:mtl}. Multi-task learning with PFI and OS always leads to improved PFI prediction performance compared to single-task SurvivalNet trained with PFI labels only.

\begin{figure}
    \centering
    \includegraphics[width=.9\textwidth]{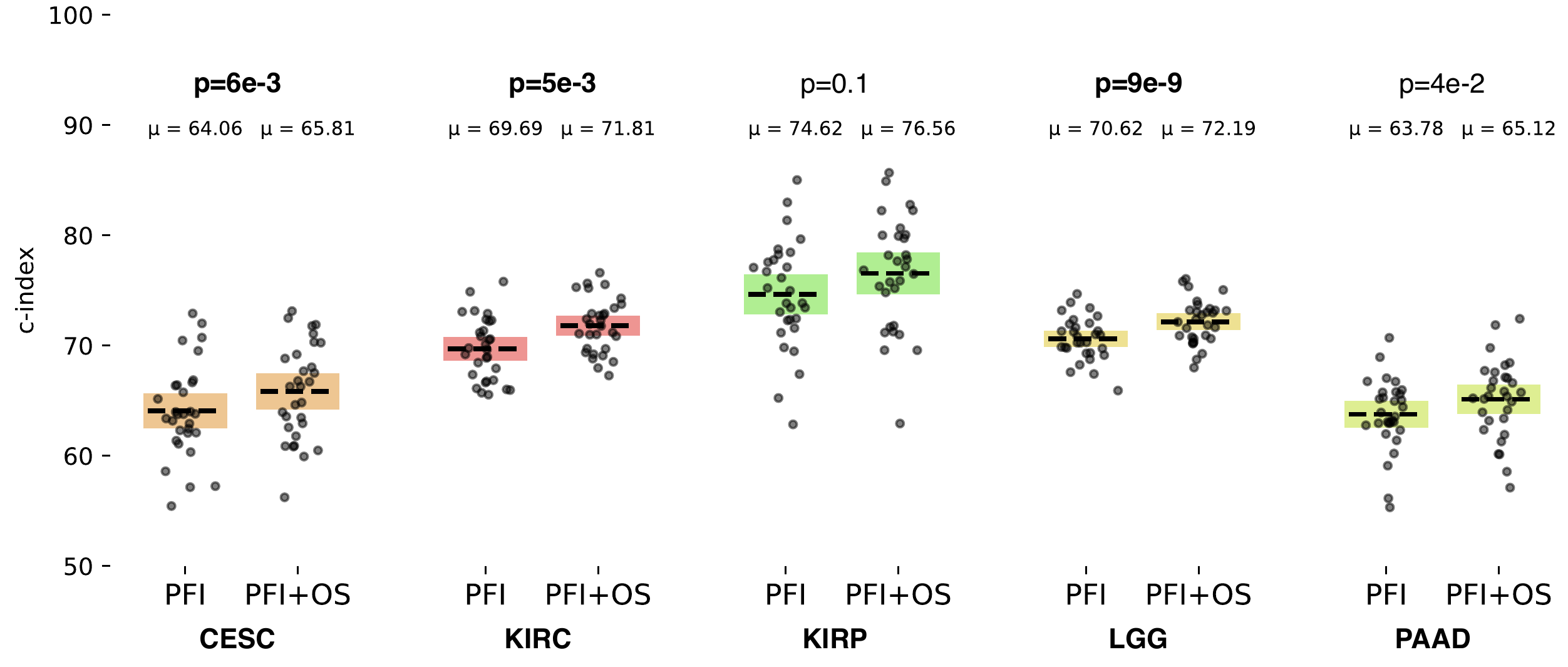}
    \caption{\footnotesize{Progression-free survival (PFI) prediction performance with and without multi-task learning with overall survival (OS). Comparison performed on five different cancer types. Significance levels are shown on the plot for each comparison.}}
    \label{fig:mtl}
\end{figure}

\begin{table}
\centering
    \begin{tabular}{LLLLL}
        \btrule{1pt}
        \footnotesize{Cancer type}& \footnotesize{PFI+OS c-index}& \footnotesize{Improvement over PFI} & \footnotesize{Censored PFI and observed OS} \\
        \hline

    CESC & 65.83   & 1.69\%                    & 5.26\%  \\
    KIRC & 76.55   & 2.12\%                    & 11.81\%  \\
    KIRP & 76.55   & 1.35\%                    & 5.19\%  \\
    LGG  & 72.15     & 1.75\%                  & 1.75\% \\
    PAAD & 65.12     & 1.34\%                  & 9.55\% \\
    \btrule{1pt}
    \end{tabular}
\caption{\footnotesize{Progression-free survival (PFI) prediction performance with and without multi-task learning with overall survival (OS). Comparison performed on five different cancer types. Percent of samples in each cohort with censored PFI and observed OS is given in the last column.}}
\label{tab:mtl}
\end{table}

\subsection{Significance of Results}
\begin{figure}
    \centering
        \begin{tabular}{cc}
        a. \footnotesize{SurvivalNet learning curve on LGG} & b. \footnotesize{SurvivalNet learning curve on BRCA} \\[6pt]
        \includegraphics[width=.44\textwidth]{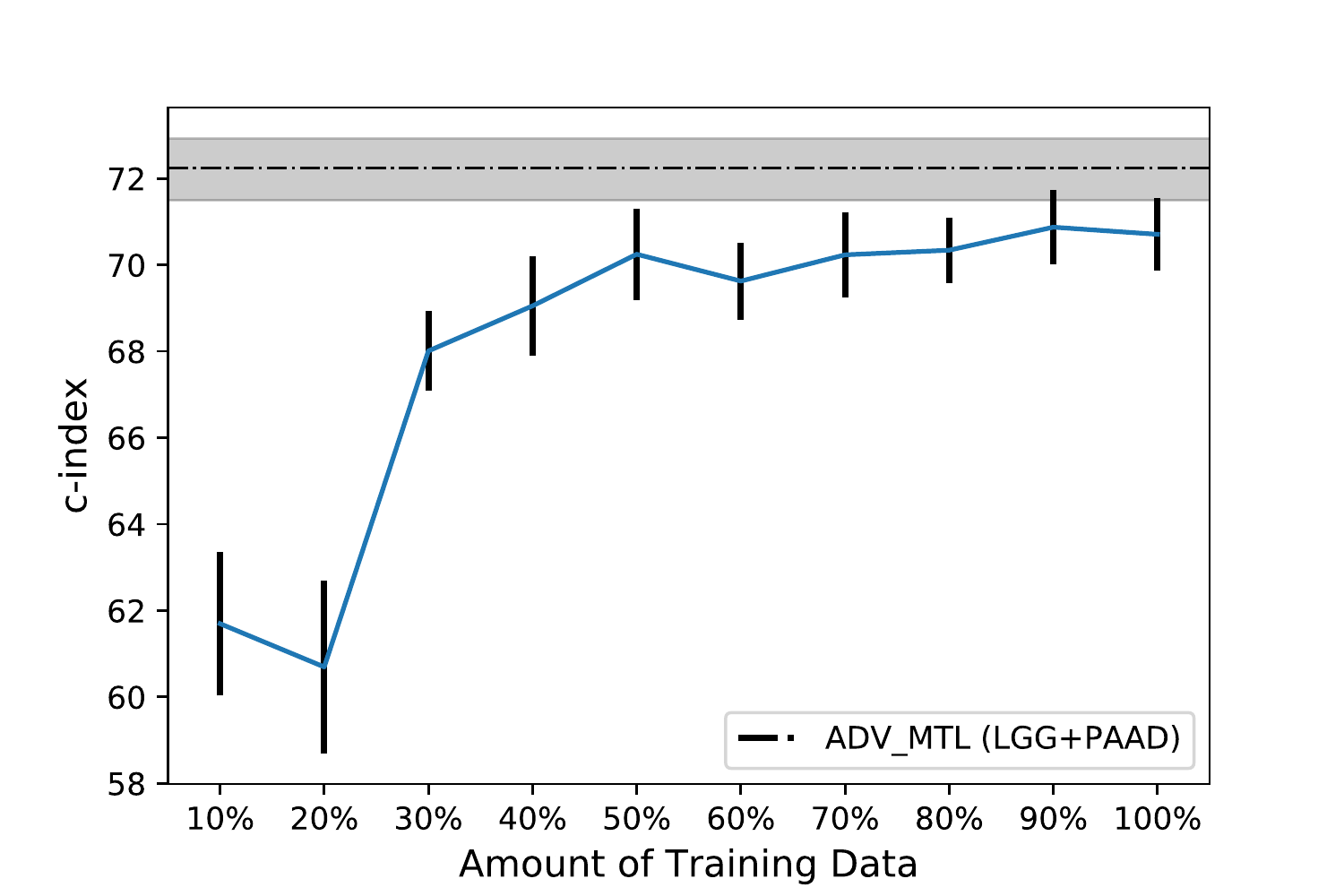} & \includegraphics[width=.45\textwidth]{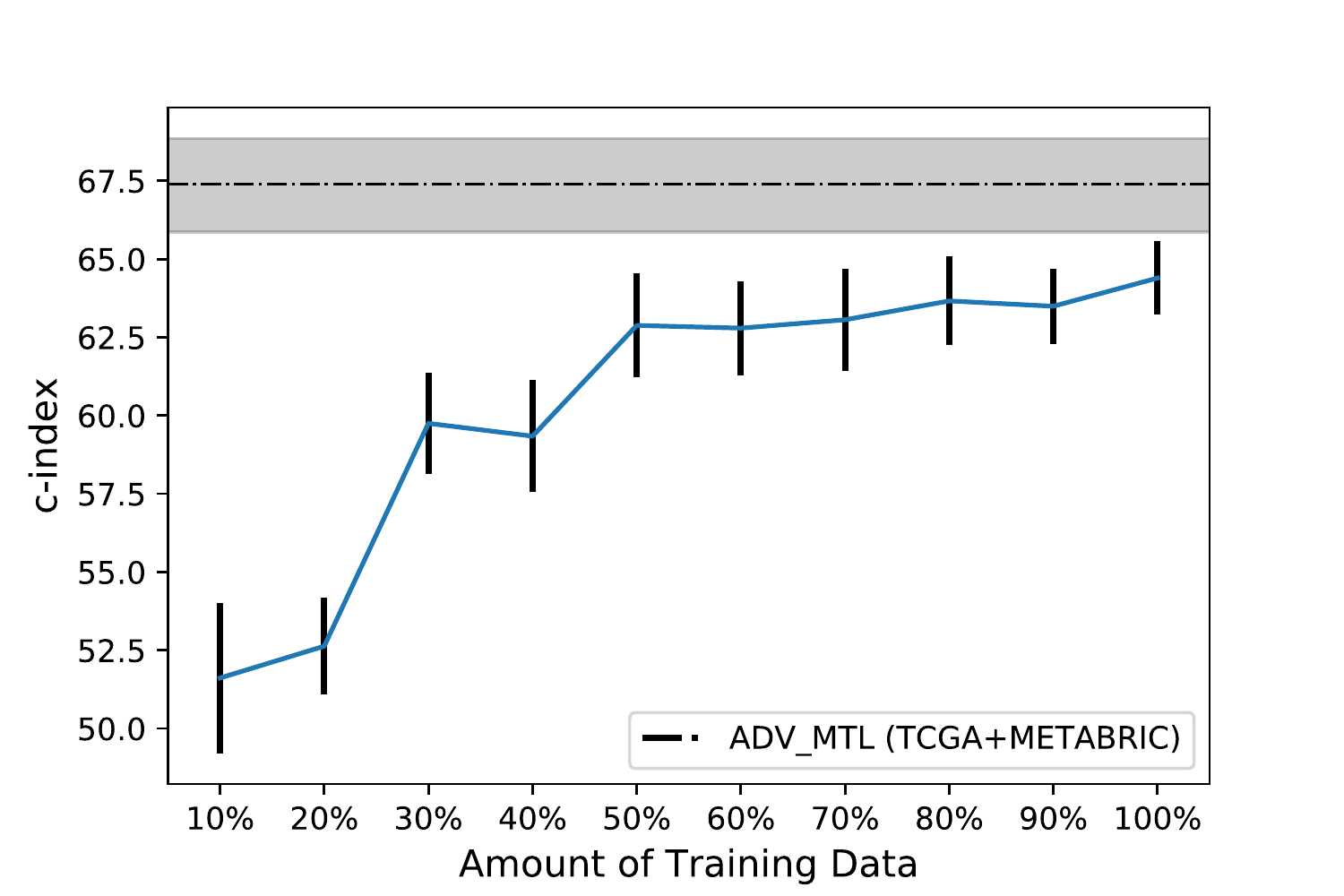}\\[6pt]        
    \end{tabular}
    \caption{\footnotesize{Learning curves of SurvivalNet on two datasets. Size of training data is gradually increased from 10\% of the original size to 100\%. The horizontal line depicts the performance of our ADV\-MTL approach. Error bars and the shaded area correspond to confidence intervals of the mean c--index.}}
    \label{fig:lc}
\end{figure}

To provide an insight into the significance of the improvement achieved by our models, we look at the learning curve of SurvivalNet on two target datasets. Learning curves were obtained by training SurvivalNet on incrementally more training samples (using the same procedure described in section \ref{subsec:training}) and testing on a fixed sized test set ($40\%$ of data, consistent with the rest of experiments). As shown in Figure \ref{fig:lc}, the performance improvement achieved by ADV-MTL over SurvivalNet trained on the full training dataset exceeds the improvement resulting from doubling the size of target training data (from 50\% to 100\% of training set). Considering the cost of obtaining labeled samples, we believe the ADV-MTL approach can save researchers resources by enabling the integration of heterogeneous cohorts in training.

\label{sec:lc}

\section{Discussion and Future Work} 

Data insufficiency is known to hinder successful application of machine learning models to high dimensional genomic data. In this paper, we study two neural network models for learning from combinations of heterogeneous datasets to tackle this issue. Significant improvement was achieved by identifying and integrating independent cohorts diagnosed with the same cancer type, and training the proposed models with the integrated data. We show how even genomic data obtained from different tumor sites can be used to augment training data and improve performance. Moreover, different outcome labels of the same cohort were used in multi-task learning to alleviate the outcome censoring issue and significant improvement was observed in most cases. 

We show that the integration of heterogeneous datasets using our proposed method is a reasonable alternative to acquisition of new training data from the target distribution which may be expensive or impossible due to practical constraints (funding, availability, clinical setting, etc).The ideal solution to any data insufficiency issue is enhanced data collection and standardization efforts. However, in settings where this is impractical, employing techniques like ADV-MTL and MTL can help address this at no extra cost. While our work focused on combining datasets from the same feature space, future work may apply or extend the proposed models to scenarios 2 and 4 introduced in Section \ref{sec:mtl4sa}, namely multi-task learning using datasets with different feature spaces and/or label spaces. Studying different cancer subtypes (eg. breast cancer histologic subtypes) under a multi-task learning setting could also lead to improved prediction.

\acks{This work was funded by National Institutes of Health National Cancer Institute U01CA220401.}

\bibliography{references.bib}
\end{document}